\documentclass[twocolumn]{aastex631}

\begin{document}
\title{The Merger Rate of Primordial Black Hole-Neutron Star Binaries in Ellipsoidal-Collapse Dark Matter Halo Models}

\author[0000-0002-6349-8489]{Saeed Fakhry} 
\email{s\_fakhry@sbu.ac.ir}
\affiliation{Department of Physics, Shahid Beheshti University, Evin, Tehran 19839, Iran}
\affiliation{PDAT Laboratory, Department of Physics, K.N. Toosi University of Technology, P.O. Box 15875-4416, Tehran, Iran}
\author{Zahra Salehnia}
\email{zahra.salehnia@email.kntu.ac.ir}
\affiliation{Department of Physics, K.N. Toosi University of Technology, P.O. Box 15875-4416, Tehran, Iran}
\affiliation{PDAT Laboratory, Department of Physics, K.N. Toosi University of Technology, P.O. Box 15875-4416, Tehran, Iran}
\author{Azin Shirmohammadi}
\email{azinshr@email.kntu.ac.ir}
\affiliation{Department of Physics, K.N. Toosi University of Technology, P.O. Box 15875-4416, Tehran, Iran}
\affiliation{PDAT Laboratory, Department of Physics, K.N. Toosi University of Technology, P.O. Box 15875-4416, Tehran, Iran}
\author{Javad T. Firouzjaee}
\email{firouzjaee@kntu.ac.ir}
\affiliation{Department of Physics, K.N. Toosi University of Technology, P.O. Box 15875-4416, Tehran, Iran}
\affiliation{School of Physics, Institute for Research in Fundamental Sciences (IPM), P.O. Box 19395-5531, Tehran, Iran}
\affiliation{PDAT Laboratory, Department of Physics, K.N. Toosi University of Technology, P.O. Box 15875-4416, Tehran, Iran}

\begin{abstract}
In this work, we calculate the merger rate of primordial black hole-neutron star (PBH-NS) binaries within the framework of ellipsoidal-collapse dark matter models and compare it with that obtained from spherical-collapse dark matter halo models. Our results exhibit that ellipsoidal-collapse dark matter halo models can potentially amplify the merger rate of PBH-NS binaries in such a way that it is very close to the range estimated by the LIGO-Virgo observations. In contrast, spherical-collapse dark matter halo models cannot justify PBH-NS merger events as consistent results with the latest gravitational wave data reported by the LIGO-Virgo collaboration. In addition, we calculate the merger rate of PBH-NS binaries as a function of PBH mass and fraction within the context of ellipsoidal-collapse dark matter halo models. The results indicate that PBH-NS merger events with masses of $(M_{\rm PBH}\le 5 M_{\odot}, M_{\rm NS}\simeq1.4 M_{\odot})$ will be consistent with the LIGO-Virgo observations if $f_{\rm PBH}\simeq 1$.
\end{abstract}

\keywords{Primordial Black Hole --- Neutron Star --- Dark Matter --- Ellipsoidal-Collapse Halo Model}

\section{Introduction} \label{sec:intro}
Gravitational waves (GWs), as a cosmological observable, have been focused on as an interesting research topic for more than a few decades. In this regard, GWs and their direct detections have provided a new framework for evaluating a large number of astrophysical and cosmological phenomena. The compact binary merger has always been one of the potential sources for the emission of GWs \cite{2022LRR....25....1M}. Fortunately, over the last few years, dozens of GW events resulting from the merger of compact binaries have been recorded by GW detectors (e.g. \cite{2016PhRvL.116f1102A, 2016PhRvL.116x1103A, 2016PhRvL.116v1101A, 2020ApJ...896L..44A, 2020PhRvL.125j1102A}). There are three general categories of compact objects participating in merger events recorded by the ground-based GW detectors: binary black holes (BBH), black hole-neutron star (BH-NS) binaries, and binary neutron stars (BNS). Among these, a majority of the recorded GWs are attributed to BBH merger events in the mass range of $(10\,\mbox{-}\,100)\, M_{\odot}$ \cite{2019PhRvX...9c1040A, 2021PhRvX..11b1053A, 2021arXiv211103606T}. Accordingly, the origin of the formation of such BHs has been the subject of a wide range of studies (e.g. \cite{2019JCAP...02..018R, 2021MNRAS.507.5224B, 2021PhRvL.126b1103N, 2021PhRvL.127o1101N, 2022PhR...955....1M}). However, this issue remains a mysterious challenge. They may have formed through stellar collapse (possibly via different channels) \cite{2021ApJ...912...98F, 2021RNAAS...5...19R}, or they may have primordial origin due to the gravitational collapse of density fluctuations in the early Universe.

As a fascinating point, GW data reported by the LIGO-Virgo collaboration are well consistent with the primordial black hole (PBH) mergers. Theoretically, PBHs are expected to form due to the non-linear peaks in primordial density fluctuations while re-entering the horizon (see, e.g. \cite{1967SvA....10..602Z, 1971MNRAS.152...75H, 1974MNRAS.168..399C}). According to a wide range of studies, for the formation of PBHs, providing a critical state in primordial density fluctuations is required, which can be realized by exceeding a threshold value (e.g. \cite{1999PhRvD..60h4002S, 2007CQGra..24.1405P, 2013CQGra..30n5009M, 2014JCAP...07..045Y, 2015arXiv150402071B, 2017JCAP...06..041A}). These conditions can yield a direct collapse of primordial density fluctuations into PBHs. In addition to their formation mechanism, PBHs can span a wide range of masses, which makes them different from astrophysical BHs \cite{2018CQGra..35f3001S}. Over the last few decades, powerful observational methods have been employed to constrain the abundance of PBHs in various mass ranges, making them reliable frameworks for studying the early Universe at small scales (e.g. \cite{2017PhRvD..96b3514C, 2018JCAP...04..007L, 2021RPPh...84k6902C}). Moreover, by assuming their participation in merger events associated with GW detectors, strong constraints on the abundance of PBHs can be obtained. Despite many theoretical uncertainties, stellar-mass PBHs (those consistent with GW observations) could have a significant contribution to the structure of dark matter \cite{2016PhRvL.116t1301B, 2017PDU....15..142C}. However, the binary PBH formation scenario in the early Universe suggests that the contribution of such BHs in dark matter structure should be very small to be consistent with the LIGO-Virgo observations \cite{2016PhRvL.117f1101S}.

As mentioned earlier, another type of GW event that can be recorded by the LIGO-Virgo detectors is the BNS merger. BNS systems can form dynamically in dense regions such as star clusters \cite{2020ApJ...888L..10Y}. Such events emit electromagnetic waves along with GWs, which provide a suitable framework for the study of multi-messenger astronomy. It should be noted that without analyzing separate electromagnetic signals and specifying sufficient upper limits on the tidal-deformability parameter of BNS, it is not easy to distinguish NS candidates from solar-mass BHs \cite{2017PhRvL.119p1101A, 2017ApJ...848L..12A}. This is one of the challenging issues related to multi-messenger astronomy, which is at the forefront of studies in this research area (e.g. \cite{2021JCAP...10..019T, 2021PhRvL.126n1105D}).

In addition to BBH and BNS mergers, the third category of GW events belongs to BH-NS binary mergers, which can also contain important information about multi-messenger observations \cite{2021FrASS...8...39R}. BH-NS binary mergers can also produce an electromagnetic signal along with the GW during the merger phase \cite{2020EPJA...56....8B}. Usually, such events involve a post-merger phase in which the residual matter from the NS is accreted by the BH and creates a luminous event \cite{2021ApJ...923L...2F}. BH-NS binary mergers are of particular interest since they can provide unique information about the NS nuclear equation of state and accretion processes of BHs, along with constraining their spin and abundance (see, e.g. \cite{2019PhRvD.100f3021H, 2019PhRvL.123d1102Z, 2021ApJ...918L..38F, 2021PhRvD.104l3024T}). Recently, the first two direct events associated with BH-NS mergers were reported by the LIGO-Virgo detectors, whose mass components have been estimated as ($8.9^{+1.2}_{-1.5}M_{\odot}, 1.9^{+0.3}_{-0.2} M_{\odot}$) and ($5.7^{+1.8}_{-2.1}M_{\odot}, 1.5^{+0.7}_{-0.3}M_{\odot}$), respectively \cite{2021ApJ...915L...5A}. Many uncertainties surround the formation and merging of BH-NS binaries. However, the evolution of the field binaries can be a viable framework to describe this class of merger events. GW detectors are expected to capture an outstanding number of BH-NS merger events in the upcoming future. In light of this, it seems essential to fully understand the formation origin of compact objects involved in these events.

In Ref.~\cite{2022ApJ...931....2S}, within the context of spherical-collapse dark matter halo models, the possibility that the BH components of BH-NS events may have a primordial origin is explored. Under such assumptions, the merger rate of PBH-NS binaries has been calculated and its results compared with the mergers estimated by the LIGO-Virgo detectors. Thus, it has been argued that BH components of such events have a non-primordial origin. However, in contrast to the formation mechanism of PBH binaries that could have formed in the early Universe, the formation scenario of PBH-NS binaries is specific to the late-time Universe, i.e., after the formation of cosmological and astrophysical structures. Therefore, more realistic dark matter halo models are expected to have a significant effect on the density and velocity distribution of the PBHs participating in such events \cite{2021PhRvD.103l3014F, 2022PhRvD.105d3525F}. As an important example, in our previous works, we have shown that spherical-collapse dark matter halo models cannot justify the recorded BBH mergers during the third observing run (O3) in the framework of the PBH scenario, while more realistic halo models (e.g., those with ellipsoidal collapse) can generate consistent PBH mergers with GW observations (see, e.g. \cite{2021PhRvD.103l3014F, 2022PhRvD.105d3525F}).

In this work, we propose to employ the ellipsoidal-collapse dark matter halo models to calculate the merger rate of PBH-NS binaries. In this respect, the outline of this work is as follows. In Sec.\,\ref{sec:ii}, we introduce a convenient dark matter halo model and discuss some quantities such as halo density profile, halo mass function, and concentration-mass-redshift relation. Then, in Sec.\,\ref{sec:iii}, we calculate the merger rate of PBH-NS binaries within the context of ellipsoidal-collapse dark matter halo models and compare it with the corresponding results obtained from spherical-collapse dark matter halo models. Moreover, we calculate the redshift evolution of the merger rate of PBH-NS binaries in the framework of ellipsoidal-collapse dark matter halo models. Also, we perform our analysis in terms of various PBH masses and discuss their constraints arising from the ellipsoidal-collapse dark matter halo models. Finally, we scrutinize the results and summarize the findings in Sec.\,\ref{sec:iv}.
\section{Dark matter halo models}
\label{sec:ii}
In this section, we briefly overview spherical and ellipsoidal-collapse dark matter halo models, which will be used as the main parameters in the calculation of the merger rate of PBH-NS binaries. In cosmological perturbation theory, dark matter halos are dynamically active structures in the nonlinear regime whose density distribution can be modeled by a radius-dependent function known as the halo density profile. Over the last few decades, halo density profiles have been developed by a vast range of analytical approaches and numerical simulations to accurately describe observational data related to the rotation curve of galaxies (e.g. \cite{1965TrAlm...5...87E,1983MNRAS.202..995J,1985MNRAS.216..273D,1990ApJ...356..359H,1993MNRAS.265..250D,1996ApJ...462..563N}). Regarding this, one of the most appropriate density profiles was presented by Navarro, Frenk, and White (NFW) \cite{1996ApJ...462..563N}, which has the following form
\begin{equation} \label{NFW}
\rho(r)=\frac{\rho_{\rm s}}{r/r_{\rm s}(1+r/r_{\rm s})^2}.
\end{equation}
In this relation, $r_{\rm s}$ is the scaled radius of halo, $\rho_{\rm s}=\rho_{\rm crit}\delta_{\rm c}$, $\rho_{\rm crit}$ is the critical density of the Universe at a given redshift $z$, and $\delta_{\rm c}$ is the linear threshold of overdensities that is related to the concentration parameter $C$ via the following formula
\begin{equation} \label{delta_c}
 \delta_{\rm c}=\frac{200}{3}\frac{C^3}{\ln(1+C)-C/(1+C)} \, .
\end{equation}
Another suitable density profile, which is obtained from analytical models, is defined as \cite{1965TrAlm...5...87E}
\begin{equation}
\rho(r)=\rho_{\rm s} \exp\bigg\{-\frac{2}{\alpha}\left[\left(\frac{r}{r_{\rm s}}\right)^{\alpha}-1\right]\bigg\},
\end{equation}
where $\alpha$ is the shape parameter. This relation is known as the Einasto density profile.

The halo concentration parameter characterizes the central density of dark matter halos and is specified by the following definition
\begin{equation} \label{r_200}
 C\equiv\frac{r_{\rm vir}}{r_{\rm s}},
\end{equation}
where $r_{\rm vir}$ is the halo virial radius, which covers a volume within which the average halo density is $200$ to $500$ times the critical density of the Universe. Studies derived from $N$-body simulations demonstrate that the concentration parameter is a decreasing function in terms of halo mass and is a function of redshift at fixed mass (e.g. \cite{2012MNRAS.423.3018P, 2014MNRAS.441.3359D, 2016MNRAS.460.1214L, 2016MNRAS.456.3068O}), which is consistent with the dynamics related to the merger tree of dark matter halos and their evolutions. Because smaller halos have already become virialized and are expected to be more concentrated than the larger ones. According to the studies mentioned above, the concentration-mass-redshift relation can be expressed by the $C(\nu)$ relation. In this description, $\nu(M, z)\equiv\delta_{\rm sc}(z)/\sigma(M, z)$ is a dimensionless parameter called the peak height, $\delta_{\rm sc}(z)=1.686(1 + z)$ is the threshold of overdensities in spherical-collapse dark matter halo models, and $\sigma(M, z)$ is the linear root-mean-square fluctuation of overdensities. As a consequence, the peak height parameter implies that the concentration parameter depends on mass and redshift. In this work, we use Eq.\,(C1) of \cite{2016MNRAS.460.1214L} as the concentration-mass-redshift relation of spherical-collapse dark matter halo models, and we employ Eqs.\,(34) and (36) of \cite{2016MNRAS.456.3068O} as the corresponding relations for ellipsoidal-collapse dark matter halo models.

Another essential parameter that can determine the properties of halos according to their mass is known as the halo mass function. Halo mass function provides a comprehensive picture of the mass distribution of dark matter halos. In the standard model of cosmology, the density contrast is defined as $\delta(x) \equiv [\rho(x)-\bar{\rho}]/\bar{\rho}$, where $\rho(x)$ is the density of overdense region at point $x$ and $\bar{\rho}$ is the mean density of background. Essentially, the halo mass function is a powerful probe to classify the mass of cosmological structures whose density contrasts exceed a threshold value, collapse, and yield halo virialization. If dark matter halos are supposed to be virialized under a spherically-symmetric collapse condition, then the threshold of overdensities, $\delta_{\rm sc}(z)$, will be approximately constant at narrow redshift changes. However, in ellipsoidal-collapse dark matter halo models, the situation is slightly different, which will be discussed later.

In addition, to categorize various fits provided for dark matter halos, a suitable definition for the differential halo mass function has been proposed as \cite{2001MNRAS.321..372J}
\begin{equation}
\frac{d n}{d M}=h(\sigma) \frac{\rho_{\mathrm{m}}}{M} \frac{d \ln \left(\sigma^{-1}\right)}{d M} ,
\end{equation}
where $M$ is the halo mass, $n(M)$ determines the number density of dark matter halos, $\rho_{\rm m}$ is the cosmological matter density, and $h(\sigma)$ is a function that is related to the geometrical conditions for overdensities at the collapse time.

Several studies have been conducted to provide an accurate halo mass function consistent with cosmological observations in this area (see, e.g. \cite{2003MNRAS.346..565R, 2006ApJ...646..881W, 2007MNRAS.374....2R, 2008ApJ...688..709T}). In this regard, Press and Schechter presented one of the most renowned analytical models, which is based on the spherical collapse of an overdense region \cite{1974ApJ...187..425P}. They presented the following function 
\begin{equation}
h_{\mathrm{ps}}(\sigma)=\sqrt{\frac{2}{\pi}} \frac{\delta_{\mathrm{sc}}}{\sigma} \exp \left(-\frac{\delta_{\mathrm{sc}}^2}{2 \sigma^2}\right),
\end{equation}
which is called the Press-Schechter (PS) halo mass function. Despite being consistent with cosmological observations across a wide range of halo masses, the PS mass function deviates from numerical simulations in some mass ranges.

Accordingly, Sheth and Tormen proposed a more realistic model to address this issue. They proposed a formalism based on dark matter halo formation under ellipsoidal-collapse conditions, where overdensities become qualified to pass from the following dynamical threshold value \cite{1999MNRAS.308..119S, 2001MNRAS.323....1S}
\begin{equation} 
\delta_{\mathrm{ec}}(\nu) \approx \delta_{\mathrm{sc}}\left(1+\alpha \nu^{-2 \beta}\right),
\end{equation}
where $\alpha=0.47$, and $\beta=0.615$. Compared to the corresponding quantity in spherical-collapse halo models, $\delta_{\rm ec}(\nu)$ can provide a better picture of hierarchical structure formation because it is defined in higher parameter space. Based on such assumptions, the Sheth-Tormen (ST) halo mass function can be obtained as
\begin{equation}
h_{\mathrm{st}}(\sigma)=A\sqrt{\frac{2 a}{\pi}} \frac{\delta_{\mathrm{sc}}}{\sigma} \exp \left(-\frac{a \delta_{\mathrm{sc}}^2}{2 \sigma^2}\right)\left[1+\left(\frac{\sigma^2}{a \delta_{\mathrm{sc}}^2}\right)^q\right] ,
\end{equation}
where $A=0.3222$, $a=0.707$, and $q=0.3$. Our argument of the dark matter halo models has now included all the necessary tools. In the following section, we will calculate the merger rate of PBH-NS binaries in ellipsoidal-collapse dark matter halo models, and compare it with that obtained from spherical-collapse dark matter halo models.
\section{The Merger Rate of PBH-NS Binaries}
\label{sec:iii}
Suppose inside an isolated dark matter halo, a PBH with mass $m_{1}$ abruptly encounters a NS with mass $m_{2}$ on a hyperbolic orbit, and their relative velocity at large separation is $v_{\rm rel}=|v_{1}-v_{2}|$. Under this assumption, $2$-body scattering implies that remarkable gravitational radiation propagates at the closest physical separation, which is known as the periastron. If the emitted gravitational energy exceeds the kinetic energy of the system, the compact objects will become gravitationally bound and form binary. This condition imposes a maximum value on the periastron as \cite{123456789p}
\begin{equation}\label{periast0}
r_{\rm mp} = \left[\frac{85 \pi}{6\sqrt{2}}\frac{G^{7/2}m_{1}m_{2}(m_{1}+m_{2})^{3/2}}{c^{5}v_{\rm rel}^{2}}\right]^{2/7},
\end{equation}
where $G$ is the gravitational constant and $c$ is the velocity of light. Besides, in the Newtonian limit, the impact parameter depends on the periastron as follows
\begin{equation}\label{impact}
b^{2}(r_{\rm p}) = \frac{2G(m_{1}+m_{2})r_{\rm p}}{v_{\rm rel}^{2}} + r_{\rm p}^{2}.
\end{equation}

Moreover, tidal forces produced by the surrounding compact objects on the binary can be neglected when the strong limits of gravitational focusing, i.e., $r_{\rm p}\ll b$, are established. Therefore, the cross-section for the binary formation can be specified by the following relation
\begin{equation}\label{crosssec}
\xi(m_1,m_2,v_{\rm rel})=\pi b^2(r_{\rm mp})\simeq\frac{2\pi G(m_1+m_2)r_{\rm mp}}{v^2_{\rm rel}}.
\end{equation}
Thus, by incorporating Eq.\,(\ref{periast0}) into Eq.\,(\ref{crosssec}), an explicit description of the cross-section for the binary formation can be obtained as
\begin{equation}
\xi \simeq 2\pi \left(\frac{85\pi}{6\sqrt{2}}\right)^{2/7}\frac{G^{2}(m_1+m_2)^{10/7}(m_1m_2)^{2/7}}{c^{10/7}v_{\rm rel}^{18/7}}.
\end{equation}
Accordingly, the binary formation rate in a single galactic halo is given by the following relation
\begin{equation}
\Gamma =4\pi \int_{0}^{r_{\rm vir}} \left(\frac{f_{\rm PBH}\rho_{\rm halo}}{m_1}\right) \left(\frac{\rho_{\rm NS}}{m_2}\right) \langle\xi v_{\rm rel}\rangle \, r^2\,dr,
\end{equation}
where $0 < f_{\rm PBH} \leq 1$ is the fraction of PBHs that determines their contribution to dark matter, $\rho_{\rm halo}$ is the halo density profile that can be considered as the NFW or the Einasto profiles, and the angle bracket exhibits an average over the PBH relative velocity distribution in the galactic halo. Also, $\rho_{\rm NS}$ is the NS density profile that we characterize by the following spherically-symmetric form
\begin{equation}\label{nsdensity}
\rho_{\rm NS}(r)= \rho^{*}_{\rm NS}\exp\left(-\frac{r}{r^{*}_{\rm NS}}\right)
\end{equation}
where $\rho^{*}_{\rm NS}$ and $r^{*}_{\rm NS}$ define as the characteristic density and radius of NSs. As can be seen from Eq.\,(\ref{nsdensity}), to determine the distribution of NSs, two quantities must be determined. First, the characteristic radius of NS, for which one can consider values in the range of $r^{*}_{\rm NS}\simeq (0.01\mbox{-}0.1)r_{\rm s}$ \cite{2022ApJ...931....2S}. Secondly, the characteristic density of NSs, $\rho^{*}_{\rm NS}$, needs to be determined by normalizing the distribution of NSs to their estimated population in an arbitrary galaxy. For this purpose, we use the time-independent form of the initial Salpeter stellar mass-function, which is described as $\chi(m_{*})\sim m_{*}^{-2.35}$. We also assume that the total number of stars in the mass range of $(8\mbox{-}20)\,M_{\odot}$ experience a supernova explosion, and their final product will be a NS. Thus, the number of NSs in a galaxy with stellar mass $M_{*}$ is specified by the following definition
\begin{equation}
n_{\rm NS}=M_{*}\int_{m_{*}^{\rm min}}^{{m_{*}^{\rm max}}}\chi(m_{*})dm_{*},
\end{equation}
in which $\chi(m_{*})m_{*}$ is normalized to unity. Note that to determine the galactic stellar mass, $M_{*}$, the stellar mass-halo mass relation, $M_{*}(M_{\rm halo})$, must be specified. To this end, assuming that a vast majority of NSs have been distributed in the central region of a galaxy, we employ the stellar mass-halo mass relation presented in \cite{2013ApJ...770...57B}. 
\begin{figure}[t] 
\includegraphics[width=0.47\textwidth]{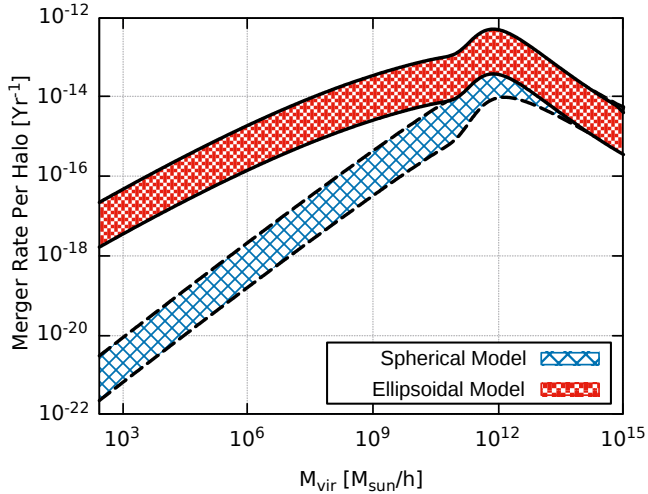}
\caption{Merger rate of PBH-NS binaries within each halo for ellipsoidal and spherical-collapse halo models as a function of halo virial mass in the present-time Universe. The shaded red band shows this relation for ellipsoidal-collapse halo models, while the shaded blue band indicates it for spherical-collapse halo models. The NFW density profile has been utilized.}
\label{fig1}
\end{figure}

\begin{table*}[t] 
\caption{Total merger event rate of PBH-NS binaries per unit time and per unit volume for a range of PBH masses, i.e., $(5\mbox{-}50)\,M_{\odot}$ while considering ellipsoidal and spherical-collapse dark matter halo models. The mass of NS is considered to be $1.4\,M_{\odot}$. The results are related to the present-time Universe and both NFW and Einasto density profiles have been considered separately.}
\centering 
\begin{tabular}{c | c | c | c }
\hline 
\hline 
$M_{\rm PBH} (M_{\odot})$ & Density Profile & Total Merger Rate~$\rm (Gpc^{-3}yr^{-1}) $ & Total Merger Rate~$\rm (Gpc^{-3}yr^{-1}) $\\ [0.5ex]
\hspace*{0.1cm}& &Spherical Model&Ellipsoidal Model\\
\hline
$5$& NFW &$1.01\times 10^{-4}-1.35\times10^{-3}$ &$0.35-4.70$\\ 
$5$& Einasto &$1.39\times 10^{-4}-1.86\times10^{-3}$ &$0.56-7.51$\\ 
\hline
$10$& NFW &$9.39\times 10^{-5}-1.25\times10^{-3}$ & $0.27-3.63$ \\ 
$10$& Einasto &$1.28\times 10^{-4}-1.71\times10^{-3}$ & $0.41-5.59$ \\ 
\hline
$20$& NFW &$8.77\times 10^{-5}-1.17\times10^{-3}$ &$0.21-2.89$\\ 
$20$& Einasto &$1.19\times 10^{-4}-1.59\times10^{-3}$ &$0.28-4.50$\\ 
\hline
$30$& NFW &$8.27\times 10^{-5}-1.10\times10^{-3}$ &$0.18-2.37$\\ 
$30$& Einasto &$1.13\times 10^{-4}-1.51\times10^{-3}$ &$0.25-3.64$\\ 
\hline
$50$ & NFW &$7.94\times 10^{-5}-1.06\times10^{-3}$ &$0.15-2.07$\\ 
$50$ & Einasto &$1.09\times 10^{-4}-1.47\times10^{-3}$ &$0.23-3.16$\\ 
\hline 
\hline
\end{tabular}
\label{table:info2}
\end{table*}

In Fig.\,\ref{fig1}, we have demonstrated the merger rate of PBH-NS binaries in each halo in the context of ellipsoidal-collapse dark matter halo models and compared it with the corresponding results obtained from spherical-collapse dark matter halo models. For this calculation, the mass of PBHs has been set as $5\,M_{\odot}$, the mass of NSs has been fixed as $1.4\,M_{\odot}$, and the NFW density profile has been employed. Also, the contribution of PBHs in dark matter is considered to be $100\%$, i.e., $f_{\rm PBH}=1$. It should be noted that the shaded regions in this figure correspond to the allowed interval, which is considered for the characteristic radius of NSs. As can be seen from the figure, the merger rate of PBH-NS binaries in the framework of ellipsoidal-collapse dark matter halo models is far higher than that obtained from spherical-collapse dark matter halo models. Additionally, for dark matter halos with smaller masses, the merger rate of PBH-NS binaries in ellipsoidal-collapse dark matter halo models has more deviations from the corresponding results extracted from spherical-collapse dark matter halo models. This can be attributed to the fact that the threshold value of overdensities under the ellipsoidal-collapse condition in halos with smaller masses has the highest deviation from that derived from the spherical-collapse condition. This result can be considered as a relative privilege of ellipsoidal-collapse dark matter halo models as more realistic frameworks in describing the evolution of dark matter halo structures. However, in our previous work, we have discussed in detail the advantage of such halo models in the proper description of BBH mergers recorded by GW detectors in the context of PBH scenario \cite{2021PhRvD.103l3014F}.

\begin{figure}[t]
\includegraphics[width=0.47\textwidth]{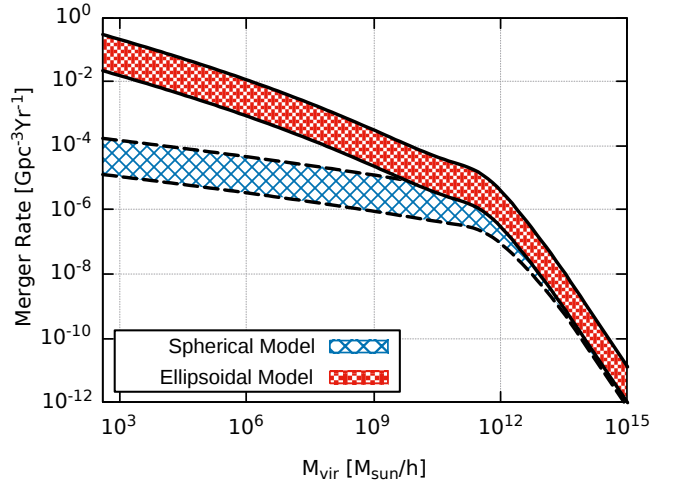}
\caption{Merger rate of PBH-NS binaries per unit time and per unit volume for ellipsoidal and spherical-collapse halo models as a function of halo virial mass in the present-time Universe. The shaded red band shows this relation for ellipsoidal-collapse halo models, while the shaded blue band represents it for spherical-collapse halo models. The NFW density profile has been utilized.}
\label{fig2}
\end{figure}

Ultimately, what can be recorded in GW detectors is an accumulation of the merger rates. In this case, by convolving the halo mass function, $dn(M)/dM$, with the rate of binary formation within each halo, $\Gamma(M)$, and integrating over a minimum mass of halos, one can calculate the total merger rate of PBH-NS binaries as follows
\begin{equation}\label{tot_mer}
 \mathcal{R}=\int_{M_{\rm min}}\frac{dn}{dM_{\rm vir}}\Gamma(M_{\rm vir})dM_{\rm vir}.
\end{equation}
In the above relation, the upper limit does not have a considerable contribution to the final result. Because the halo mass function contains an exponentially decreasing term in such a way that the merger rate changes inversely with the halo mass, while the lower limit plays a crucial role. This is consistent with the dynamics of the evolution of hierarchical structures because the smallest halos are expected to be denser than host halos. However, dynamical relaxation processes result in the evaporation of the smallest halos via the ejection of objects. On the other hand, compensating factors for evaporation, such as merger and accretion, reach their lowest level of effectiveness during the dark energy-dominated era. Thus, dynamical relaxation processes could have led to the complete evaporation of the smallest halos during the recent $3$ Gyr in such a way that the merger signals from some halos can be ignored \cite{2016PhRvL.116t1301B}. Therefore, the evaporation time of dark matter halos can guarantee their survival in the present-time Universe. In our previous works, we have shown that the evaporation time of dark matter halos with a typical mass of $400\,M_{\odot}$, which contain PBHs with a mass of $30\,M_{\odot}$, is about $3$ Gyr (see, \cite{2021PhRvD.103l3014F, 2022PhRvD.105d3525F} for more details). Hence, dark matter halos with a typical mass of $400\,M_{\odot}$ can be considered as the minimum mass in our analysis. However, dark matter halos containing PBHs lighter than $30\,M_{\odot}$ can have masses slightly smaller than $400\,M_{\odot}$ to survive in the present-time Universe, while dark matter halos containing PBHs heavier than $30\,M_{\odot}$ can have masses a bit greater than $400\,M_{\odot}$ to fulfill the corresponding conditions.

In Fig.\,\ref{fig2}, we have shown the merger rate of PBH-NS binaries per unit volume and time in the framework of ellipsoidal-collapse dark matter halo models for the NFW density profile and compared it with the corresponding findings obtained from spherical-collapse dark matter halo models. To perform these calculations, we have used the PS halo mass function and the Ludlow concentration-mass-redshift relation for spherical-collapse dark matter halo models, while for ellipsoidal-collapse dark matter halo models we have employed the ST mass function and the Okoli-Afshordi concentration-mass-redshift relation. As is clear from the figure, compared to the results obtained from spherical-collapse dark matter halo models, ellipsoidal-collapse dark matter halo models can sufficiently amplify the merger rate of PBH-NS binaries. By integrating over the surface below the curves, one can obtain the total merger rate of PBH-NS binaries. Also, it can be inferred from the figure that the merger rate of PBH-NS binaries is inversely proportional to the minimum mass of dark matter halos, which is consistent with the argument presented earlier. In other words, the results confirm the dynamics associated with hierarchical structures in such a way that the density of dark matter particles changes inversely with the mass of halos.

Up to here, we have only shown the results for the NFW density profile. In Table~\ref{table:info2}, we have also provided the merger rate of PBH-NS binaries for both mentioned halo models in terms of a range of PBH masses, i.e., $M_{\rm PBH}=(5\mbox{-}50)M_{\odot}$ while taking into account NFW and Einasto density profiles. In addition, for the Einasto density profile, we have set the value of the shape parameter prescribed in \cite{1965TrAlm...5...87E}. It can be inferred from the Table that the merger rate of PBH-NS binaries changes inversely with the mass of PBHs, which seems reasonable. Because the number density of PBHs decreases with increasing their masses. Consequently, smaller PBHs are more likely to participate in PBH-NS binary formations than larger ones. Moreover, it can be deduced from the results that the merger rate of PBH-NS binaries for the Einasto density profile is slightly higher than that obtained from the NFW density profile. 

\begin{figure}[t!]
\includegraphics[width=0.47\textwidth]{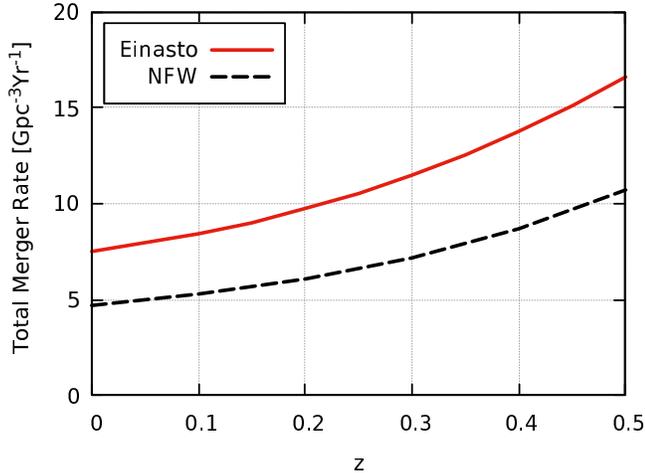}
\caption{Upper limit of the merger event rate of PBH-NS binaries for ellipsoidal-collapse dark matter halo models as a function of redshift. The solid (red) line represents this relation for the Einasto density profile, while the dashed (black) line shows it for the NFW density profile.}
\label{fig3}
\end{figure}

In addition, the LIGO-Virgo detectors are capable of detecting compact binary mergers up to $z\simeq 0.75$ in their latest configuration update. Therefore, it is worth predicting the redshift evolution of the merger rate of PBH-NS binaries within the context of ellipsoidal-collapse dark matter halo models. Fortunately, Eq.\,(\ref{tot_mer}) is a redshift-dependent function through the concentration-mass-redshift relation and the halo mass function. Based on this, in Fig.\,\ref{fig3}, we have illustrated the redshift evolution of the merger rate of PBH-NS binaries in the framework of ellipsoidal-collapse dark matter halo models while considering NFW and Einasto density profiles. The results indicate that the merger rate of PBH-NS binaries is directly proportional to redshift changes. This means that PBHs at higher redshifts have been more likely to encounter NSs than in the present-time Universe. This can be attributed to the dynamical evolution of the dark matter distribution, which occurs over time during the processes of merger and evaporation of galactic halos. It should be noted that these results are the upper limit allowed by the appropriate interval for the characteristic radius of NSs. Therefore, the exact value of the characteristic radius of NSs remains as a theoretical uncertainty in the obtained results.

\begin{figure}[t!]
\includegraphics[width=0.45\textwidth]{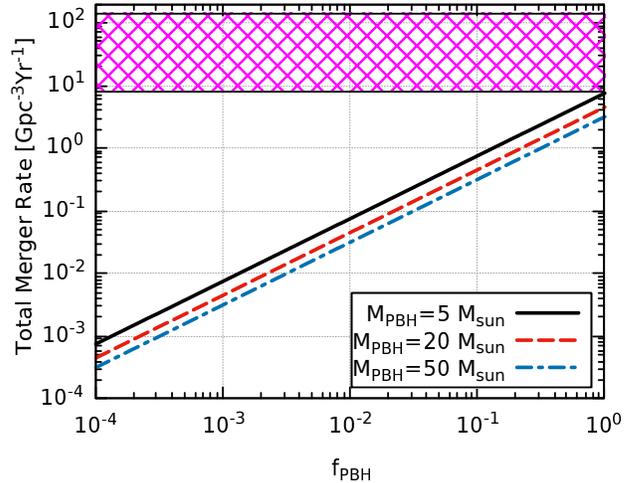}
\caption{Total merger event rate of PBH-NS binaries for ellipsoidal-collapse dark matter halo models as a function of the upper bounds on PBH fraction and their masses. The solid (black), dashed (red), and dot-dashed (blue) lines indicate this relation while considering PBH mass to be $M_{\rm PBH} = 5, 20$, and $50 M_{\odot}$, respectively. The shaded band indicates the total merger rate of BH-NS binaries recorded by the LIGO-Virgo detectors, i.e., $(7.8\mbox{-}140)\,{\rm Gpc^{-3}yr^{-1}}$.}
\label{fig4}
\end{figure}

Finally, in Fig.\,\ref{fig4}, we have depicted the merger rate of PBH-NS binaries in the framework of ellipsoidal-collapse dark matter halo models in terms of the upper bounds on the fraction of PBHs for several PBH masses. Furthermore, the shaded band shows the total merger rate of BH-NS binaries recorded by the LIGO-Virgo detectors, i.e., $(7.8\mbox{-}140)\,{\rm Gpc^{-3}yr^{-1}}$ \cite{2021arXiv211103634T}. The results demonstrate that, despite many theoretical uncertainties, PBH-NS merger events with masses of $(M_{\rm PBH}\simeq 5\,M_{\odot}, M_{\rm NS}\simeq1.4\,M_{\odot})$ will be comparable with the latest BH-NS mergers estimated by GW detectors if the fraction of PBHs is $f_{\rm PBH} \simeq 1$. While the corresponding results obtained from spherical-collapse dark matter halo models are not able to do this. Of course, it should not be overlooked that the obtained constraint on the abundance of PBHs is the upper limit allowed by the LIGO-Virgo detectors. Because the mass range of BHs participating in such events has a very strong overlap with the mass range of BHs that have resulted from stellar collapse. Therefore, the LIGO-Virgo detectors may have recorded events that included astrophysical BHs. Also, it can be inferred that to have at least one $(M_{\rm PBH}\simeq 5\,M_{\odot}, M_{\rm NS}\simeq1.4\,M_{\odot})$ event in the comoving volume $1\,{\rm Gpc^{3}}$ annually, the fraction of PBHs should be $f_{\rm PBH}\ge \mathcal{O}(10^{-1})$. This can be considered as a new constraint on the abundance of PBHs derived from the merger rate of PBH-NS binaries in ellipsoidal-collapse dark matter halo models, which is obtained for the first time via this method. 

\section{Conclusions}
\label{sec:iv}
PBHs are a special type of BHs that may have formed through the direct collapse of density fluctuations in the early Universe. PBHs are considered as a potential macroscopic candidate for dark matter. Due to their random distribution in the late-time Universe, PBHs can encounter other compact objects in dark matter halos. In contrast to the formation time of PBHs, NSs are known as another class of compact objects that are the product of a stellar collapse and supernova explosion in the late-time Universe. Over the last few years, the LIGO-Virgo collaborations have speculated about recording GWs caused by BH-NS binary mergers until they announced the first two such mergers. These merger events are important because they emit electromagnetic signals along with the propagation of GWs, leading to new possibilities in multi-messenger astronomy. In BH-NS events, the study of the origin of the BH component can contain vital information about the formation and evolution of compact objects. There is a possibility that these BHs are the final product of stellar collapse (possibly through various channels). They may also be PBHs, whose formation channels date back to the early Universe.

In this work, we have calculated the merger rate of PBH-NS binaries in the ellipsoidal-collapse of dark matter halo models and compared it with the corresponding results obtained from spherical-collapse dark matter halo models. To do this task, we have initially defined a suitable framework for dark matter halo models and introduced their key parameters, including the density profile, the concentration-mass-redshift relation, and the halo mass function. Regarding this, we have argued that the PS halo mass function and the Ludlow concentration-mass-redshift relation are sufficient for spherical-collapse dark matter halo models, whereas the ST halo mass function and the Okoli-Afshordi concentration-mass-redshift relation are employed for ellipsoidal-collapse dark matter halo models.

Having all the necessary tools for the halo modeling, we have calculated the merger rate of PBH-NS binaries within each dark matter halo. To perform this, we have set the mass of PBHs to be $5\,M_{\odot}$ and placed the mass of NSs as $1.4\,M_{\odot}$. The results indicate that the merger rate of PBH-NS binaries in the context of ellipsoidal-collapse dark matter halo models is far higher than that extracted from spherical-collapse dark matter halo models. Moreover, it has been confirmed that for halos with smaller masses, the merger rate of PBH-NS binaries in ellipsoidal-collapse dark matter halo models has more deviations from that obtained for spherical-collapse dark matter halo models. This can be referred to the fact that the threshold value of overdensities for ellipsoidal-collapse halos with smaller mass has the highest deviation from that obtained in spherical-collapse halo models.

Furthermore, we have calculated the accumulated merger rate of PBH-NS binaries in ellipsoidal-collapse dark matter halo models in the present-time Universe and compared it with the corresponding findings obtained from spherical-collapse dark matter halo models. According to the results, ellipsoidal-collapse dark matter halo models can potentially generate a sufficient merger rate of PBH-NS binaries. In contrast, spherical-collapse dark matter halo models cannot perform this task. We have also provided the merger rate of PBH-NS binaries for both halo models in terms of a range of PBH masses, i.e., $M_{\rm PBH}=(5\mbox{-}50)M_{\odot}$ while considering NFW and Einasto profiles. The results indicate that the merger rate of PBH-NS binaries for the Einasto density profile is slightly higher than that derived from the NFW density profile. Moreover, it has been observed that the merger rate of PBH-NS binaries is inversely proportional to the mass of PBHs. Consequently, smaller PBHs are more likely to involve in PBH-NS binary formations than larger ones.

We have also calculated the redshift evolution of the merger rate of PBH-NS binaries in ellipsoidal-collapse dark matter halo models while taking into account NFW and Enasto density profiles. The results show that the merger rate of PBBH-NS binaries varies directly with redshift changes. This means that PBHs at higher redshifts have been more likely to encounter NSs than in the present-time Universe. This can be explained by the dynamical evolution of the dark matter distribution, which happens through the merger and evaporation of galactic halos. 

Finally, we have calculated the merger rate of PBH-NS binaries within the context of ellipsoidal-collapse dark matter halo models as a function of the upper bounds on the fraction of PBHs for several PBH masses. It has been inferred that ellipsoidal-collapse dark matter halo models will be able to justify the merger rate of PBH-NS binaries with masses of $(M_{\rm PBH}\le 5\,M_{\odot}, M_{\rm NS}\simeq1.4\,M_{\odot})$ in the framework of LIGO-Virgo sensitivity if $f_{\rm PBH} \simeq 1$. In contrast, similar results cannot be extracted from spherical-collapse dark matter halo models. This result reinforces the argument that the LIGO-Virgo detectors will be more likely to detect astrophysical BH-NS binary mergers if spherical-collapse dark matter halo models are credible. Moreover, the findings indicate that to have at least one $(M_{\rm PBH}\simeq 5\,M_{\odot}, M_{\rm NS}\simeq1.4\,M_{\odot})$ event in the comoving volume $1\,{\rm Gpc^{3}}$ annually, the fraction of PBHs is constrained as $f_{\rm PBH}\ge \mathcal{O}(10^{-1})$. Interestingly, despite many theoretical uncertainties in the exact justification of GW observations, this new constraint on the abundance of PBHs was obtained via the calculation of the merger rate of PBH-NS binaries in ellipsoidal-collapse dark matter halo models.

In this work, we have tried to show that ellipsoidal-collapse dark matter halo models for calculating the merger rate of PBH-NS binaries can be a suitable framework for justifying BH-NS mergers recorded by GW detectors. However, there are many theoretical uncertainties in the presented model, which may not fulfill the ideal conditions for its realization. For instance, the mass interval considered for PBHs has a strong overlap with the allowed mass interval for astrophysical BHs. Consequently, it is still possible that the BH components contributing to BH-NS binary events are of astrophysical origin. Also, our argument was based on a two-body interaction leading to the formation of BH-NS binaries under a strong limit of gravitational focusing. Therefore, tidal effects caused by the surrounding compact objects on the formed binaries are not included in our analysis and remain an uncertainty in the presented model. Additionally, the exact value of the characteristic radius of NSs should be considered as another theoretical uncertainty. It is hoped that with the further development of instruments in the upcoming future, one can better understand the merger events of compact binaries.

\section*{Acknowledgments}
\noindent
We would like to gratefully acknowledge the anonymous reviewer for the insightful suggestions and comments.

\bibliography{sample631}{}
\bibliographystyle{aasjournal}

\end{document}